\documentclass[usenatbib]{mnras}
\usepackage[T1]{fontenc}
\usepackage{ae,aecompl}
\usepackage{graphicx}
\usepackage{amsmath}
\usepackage{amssymb}

\def\simgreat{\lower2pt\hbox{$\buildrel {\scriptstyle >}
   \over {\scriptstyle\sim}$}}
\def\simless{\lower2pt\hbox{$\buildrel {\scriptstyle <}
   \over {\scriptstyle\sim}$}}
\def\msun{\,{\rm M_\odot}}
\def\Msun{\,{\rm M_\odot}}

\title[The lifetime of binary BHs in S\'ersic galaxy models]{The lifetime of binary black holes in S\'ersic galaxy models}

\pubyear{2019}

\begin{document}

\author[N. Biava et al.]{Nadia Biava,$^1$ Monica Colpi,$^{1,2}$ Pedro R.~Capelo,$^3$ Matteo Bonetti,$^{2,4}$\newauthor Marta Volonteri,$^5$
Tomas Tamfal,$^3$ Lucio Mayer,$^3$  and Alberto Sesana$^{1,6}$\\
$^1$Department of Physics G. Occhialini, University of Milano-Bicocca, Piazza della Scienza 3, 20126 Milano, Italy\\
$^2$National Institute of Nuclear Physics INFN, Milano-Bicocca, Piazza della Scienza 3, 20126 Milano, Italy\\
$^3$Center for Theoretical Astrophysics and Cosmology, Institute for Computational Science, University of Zurich,\\
Winterthurerstrasse 190, CH-8057 Z{\"u}rich, Switzerland\\
$^4$DiSAT, Universit\`{a} degli Studi dell'Insubria, Via Valleggio 11, 22100 Como, Italy\\
$^5$Sorbonne Universit\'{e}, CNRS, UMR 7095, Institut d'Astrophysique de Paris, 98 bis bd Arago, 75014 Paris, France\\
$^6$School of Physics and Astronomy and Institute of Gravitational Wave Astronomy,\\ University of Birmingham, Edgbaston B15 2TT, UK}

\label{firstpage}
\pagerange{\pageref{firstpage}--\pageref{lastpage}}
\maketitle

\begin{abstract}
In the local universe, black holes of $10^{5-6}\msun$ are hosted in galaxies displaying a variety of stellar profiles and morphologies. These black holes are the anticipated targets of LISA, the Laser Interferometer Space Antenna that will detect the low-frequency  gravitational-wave signal emitted by binary black holes in this mass interval. In this paper, we infer upper limits on the lifetime of binary black holes of  $10^{5-6}\msun$ and up to $10^8\msun$, forming in galaxy mergers, exploring two underlying stellar density profiles, by Dehnen and by  Prugniel \& Simien, and by exploiting local scaling relations between the mass of the black holes and several quantities of their hosts. We focus on the phase of the dynamical evolution when the binary is transitioning from the hardening phase ruled by the interaction with single stars to the phase driven by the emission of gravitational waves. We find that different stellar profiles predict very distinct trends with binary mass, with lifetimes ranging between fractions of a Gyr to more than 10~Gyr, and with a spread of about one order of magnitude, given by the uncertainties in the observed correlations, which are larger in the low-mass tail of the observed black hole population.\\

\end{abstract}


%%%%%%%%%%%
%Section 1%%%%%
%%%%%%%%%%%

\section{Introduction}\label{sec:introduction}

The formation of a binary of two massive black holes in the aftermath of a galaxy-galaxy collision and its hardening down to the
gravitational-wave (GW) domain is a challenging problem in stellar dynamics. 
Detailing the black hole dynamics from the large scale of a cosmological galactic merger (hundreds of kpc) to the 
tiny scale of GW inspiral (of the order of a few $\mu$pc) is instrumental in predicting the rate of black hole coalescences
expected to occur during the hierarchical assembly of galaxies \citep{Enoki05,Rhook05,Sesana2011rate,Plowman11,Klein2016,Tamanini16,Bonetti2018b,Dayal2018,Ricarte18}. The environment in which black hole pairing, binary formation, and contraction occur varies immensely. 
After years of studies, we learned that gas-free and gas-rich galaxy mergers show both extremely complex and non-universal black hole dynamics \citep{Colpi2014, Mayer2013}.

There exist three main phases that black holes experience along the path to coalescence: (I) the early phase of pairing 
under dynamical friction in the stellar bulge of the post-merger galaxy, ending with the formation of a close Keplerian binary; (II) the phase of hardening 
by close encounters with single stars plunging on nearly radial orbits on to the binary or by viscous/gravitational angular momentum transport with gas in a circumbinary disc;  and finally (III) the phase of GW inspiral.

With advances in numerical simulations, the duration of each phase is modulated by complex, interrelated processes.
The inclusion of 
cosmology-driven initial conditions, cosmic gas inflows, rich galaxy morphology, asymmetries in the stellar distributions, gas dynamics, and a wider spectrum of galaxy and black hole mass ratios reveals the occurrence of failures, bottlenecks, delays, rapid sinking, and/or erratic dynamics 
\citep{Begelman1980, MilosavljevicMerritt2003, Berczik2006, Mayer2007, Dotti2007, Callegari2008, Lodato2009, Callegari2011, Khan2011, Sesana2011, Khan2013, Fiacconi2013, Vasiliev2014, Capelo2015, Roskar2015, delValle2015, Lupi2015, Mayer2016, Goicovic2017, CapeloDotti2017, SouzaLima2017, Pfister2017, Tamburello2017, Tremmel2017, Bonetti2018a}. 

In this paper, we study the stellar hardening of binary black holes in the mass interval $10^{5-8}\msun.$ These binaries are key targets for LISA, the Laser Interferometer Space Antenna \citep{Amaro-Seoane2017,Barack2018} that will primarily explore the low-mass tail of the supermassive black hole mass distribution. These black holes are now being discovered in nearby dwarf galaxies  through their electromagnetic emission \citep{Reines2013, Sartori2015, Pardo2016, Mezcua2017}.

As black hole coalescences of $10^{4-7}\msun$ can be detected with LISA from redshift $z\sim 0$ up to $z \sim 15$--20, when seed black holes form \citep{Latif2016},
the environment in which they form and couple at dynamical level varies and evolves with redshift at the rhythm of the cosmic assembly of structures.
High-resolution zoom-in simulations indicate that the stellar component, even in a gas-rich galaxy merger, is instrumental in driving the black holes to coalescence. 
A dense stellar cusp surrounding one or both black holes may form following a merger-driven, central gas inflow \citep{VanWassenhove2014,Khan2016}, and it is in this
dense environment that dynamical friction and later scattering of stars in a triaxial remnant guide the binary contraction. 

Very recently, \cite{Tamfal2018} investigated the early phase of pairing of two black holes of $10^5\msun$ in a merger between two
dark-matter-dominated dwarf galaxies, using collisionless $N$-body simulations. They found that the efficiency of pairing, prior to binary formation and stellar hardening, 
depends sensibly on the steepness of the central dark matter density profile. Only cuspy dark matter profiles appear to be favourable to forming a Keplerian binary (of mean eccentricity $e \sim 0.5$), i.e. a bound state on sub-pc scales, when the mass in stars and dark matter inside the orbit decreases below the sum of the black hole masses. Due to the limited resolution ($\sim$1 pc) of these simulations, the dynamics of the black holes could not be reliably followed down to the stages in which close encounters with single stars and GW emission become important.

In this paper, we explore this next stage 
using analytical tools in order to guide future simulations
on the dynamical evolution of black holes in dwarfs.
We aim at determining the binary lifetime in its hardening phase over a mass range that extends from $10^8\msun$ down to the least massive supermassive black holes of $10^5\msun$.

Direct $N$-body simulations of
black hole hardening in a triaxial stellar background indicate that 
a departure from spherical symmetry in the stellar nucleus of the galaxy (after the merger) keeps the rate of interaction of stars with the binary at a high enough level,
even when the system is collisionless, so that the binary continues to shrink rather rapidly. A sufficiently high number of stars moving on centrophilic orbits appears to be  present in non-spherical potentials to let the binary enter the GW driven phase \citep{Merritt2004, Holley2006}. Using a Monte Carlo method, \cite{Vasiliev2015} confirmed this trend in triaxial galaxies  without relaxation and showed that even a moderate departure from axisymmetry is sufficient to keep the binary shrinking \citep{Vasiliev2016}.
After performing a detailed comparison between direct $N$-body simulations and a hybrid model based on three-body scattering experiments \citep[see, e.g.][]{Quinlan1996,Sesana2006},
\cite{SesanaKhan2015} conjectured that the ($N$-body) binary hardening rate is equivalent to that of a binary embedded in a field of stars 
inside a spherical potential where key quantities such as the stellar density and velocity dispersion are evaluated at the binary gravitational sphere of influence.
This enables to estimate the lifetime of the binary forming in a merger, based solely on a few key parameters of the host.
\cite{SesanaKhan2015} confined their analysis to galaxies described by a \citet{Dehnen1993} stellar  profile and black holes with masses in excess of $10^7\msun,$ finding hardening time-scales ranging from a fraction of a Gyr to $\sim$10~Gyr, with a strong dependence on the binary eccentricity.
In light of recent findings that attribute to dwarf galaxies a variety of morphologies, the choice of the stellar density and velocity dispersion profile becomes then critical. 

\citeauthor{Reines2013} (\citeyear{Reines2013}; hereafter RGG) assembled the largest sample of local dwarf galaxies (with stellar masses in the interval $10^{8.5} < M_*<10^{9.5}\msun$) hosting a nuclear black hole with median mass of $2\times 10^5\msun$, detected as a low-luminosity active galactic nucleus (see \citealt{Mezcua2018} for a sample at high redshift). These galaxies display different morphologies, 
presence or absence of central pseudo-bulges, and different levels of compactness, with \citet{Sersic1963,Sersic1968} indices varying from 0.8 to 6. As an example, the dwarf disc galaxy RGG~118 hosts the least massive black hole ever detected 
with mass  $\sim$$5\times 10^4\msun$. 
It contains a pseudo-bulge of $10^{8.5}\msun$, fit by an inner S{\'e}rsic profile of index $n=0.8\pm 0.1$, placing the black hole mass below the extrapolated value from the $M_{\rm BH}$--$M_{\rm sph,*}$ relation
defined by elliptical/S0 more massive galaxies \citep{Baldassare2015, Baldassare2017}, where $M_{\rm sph,*}$ is the total mass of the stellar spheroidal.   
The departure from the $M_{\rm BH}$--$M_{\rm sph,*}$ relation is observed 
in a much wider sample of dwarf and spiral mega-maser galaxies \citep{KormendyHo2013, Lasker2016, ReinesVolonteri2015, Greene2008}.

\cite{Savorgnan2013} have highlighted the occurrence of a correlation between the black hole mass and the S\'ersic index of the host, showing that less massive black holes (below $10^7\msun$ and of interest for LISA) 
live preferentially in galaxies with low values of $n<2$.  
To anchor the black hole mass and, by extrapolation, the mass of the binary to the stellar profile of the underlying host, we here use the 
$M_{\rm BH}$--$M_{\rm sph, *}$ correlation, and the looser relation $M_{\rm BH}$--$n,$
as inferred in \citet{KormendyHo2013}, \citet{ScottGraham2013}, \citet{Savorgnan2013}, and \citet{Davis18}.
We  recall that the S\'ersic profile describes the surface brightness of a galaxy, and the index $n$ controls its degree of curvature,
representing the generalization of the de Vaucoulers' law \citep[$n=4$;][]{deVaucouleurs1948, deVaucouleurs1959}. The smaller is the value of $n$, the shallower and less concentrated the profile. 
Most galaxies are fit by S\'ersic profiles with indices in the range $1/2<n<10$ such that brighter, more massive galaxies tend to be fit with larger $n$.

In this paper, we provide an estimate of the black hole binary hardening times relevant for LISA considering either a Dehnen profile \citep[for a comparison with the work of ][]{SesanaKhan2015} 
and a \cite{PrugnielSimien1997} profile, the latter 
relating the stellar distribution to the  S\'ersic surface brightness profile. We also adopt values of scaling relations that correlate the black hole  mass with the stellar mass, the stellar velocity dispersion, and the S\'ersic index. We warn that, due to the large scatter in the cited relations present at the low masses explored and to the non universality of the stellar profiles that characterise these galaxies, our analysis provides median estimates of the black hole lifetimes.

We further note that 
two competing mechanisms, not included in our treatment, can affect our estimates. In galaxies flattened by rotation (induced by a merger), coalescence times 
are found to be shorter than in non-rotating galaxies  \citep{Holley-Bockelmann_Khan_2015}. Additional inputs in the modelling, such as counter-rotation, inclination of the binary orbital plane relative to the galaxy rotation, and initial eccentricity lead to shorter-lived binaries \citep{Mirza17}.
By contrast, binary hardening is in general conducive to core scouring \citep{Merritt_2006}. This process turns a power-law stellar profile into a cored profile \citep{Merritt_2006} if the time-scale for loss-cone refilling is long compared to the lifetime of the galaxy.
This is particularly relevant in massive galaxies with long relaxation time-scales \citep{Merritt_2006,KhanPreto2012}.  We recall that the mass deficit in stars caused by binary hardening is of the order of 50 per cent of the total mass of the binary \citep{Merritt_2006}.  Thus, scouring has a tangible effect in large elliptical galaxies for which the black hole-to-stellar mass ratio is large \citep{KormendyHo2013}. At lower masses, the steepening of the `black hole mass--stellar mass' relation  may reduce the effect of stellar  scouring, due to the larger  stellar mass content compared to the black hole mass. Letting the power index of the Dehnen profile  and S\'ersic index vary in our model, it is a way to capture all these uncertainties.

The paper is organized as follows. In Section~\ref{sec:binary_lifetime}, we model the transition between the hardening and GW emission as in \cite{SesanaKhan2015}, in order to
define the longest time of residence of a black hole binary in a given stellar background, which we refer to as binary lifetime.
We then introduce the \cite{Dehnen1993} and \cite{PrugnielSimien1997} models (Sections~\ref{sec:dehnen_density_profile} and \ref{sec:prugniel_and_simien_density_profile}, respectively)
and calculate characteristic time-scales and black hole binary separations for different values of the mass ratio and binary eccentricity, although in the rest of the paper we adopt circular binaries to focus on the stellar density at the binary black hole sphere of
influence. We then consider the data sample of \cite{GrahamScott2015}, and a study by \citet{Nguyen2018} to link these time-scales to the dormant black holes that we observe in the near universe. In Section~\ref{sec:Nguyen}, we show the strong dependence of our results on the choice of scaling relations, whereas a comparison between the two models is reported in Section~\ref{sec:model_comparison}. In Section~\ref{sec:conclusions}, we derive our conclusions.

%%%%%%%%%%%
%Section 2%%%%%
%%%%%%%%%%%

%%%%% FIGURE 1 %%%%%
\begin{figure*}
\centering
\vspace{-0pt}
\centering
\includegraphics[scale=0.35]{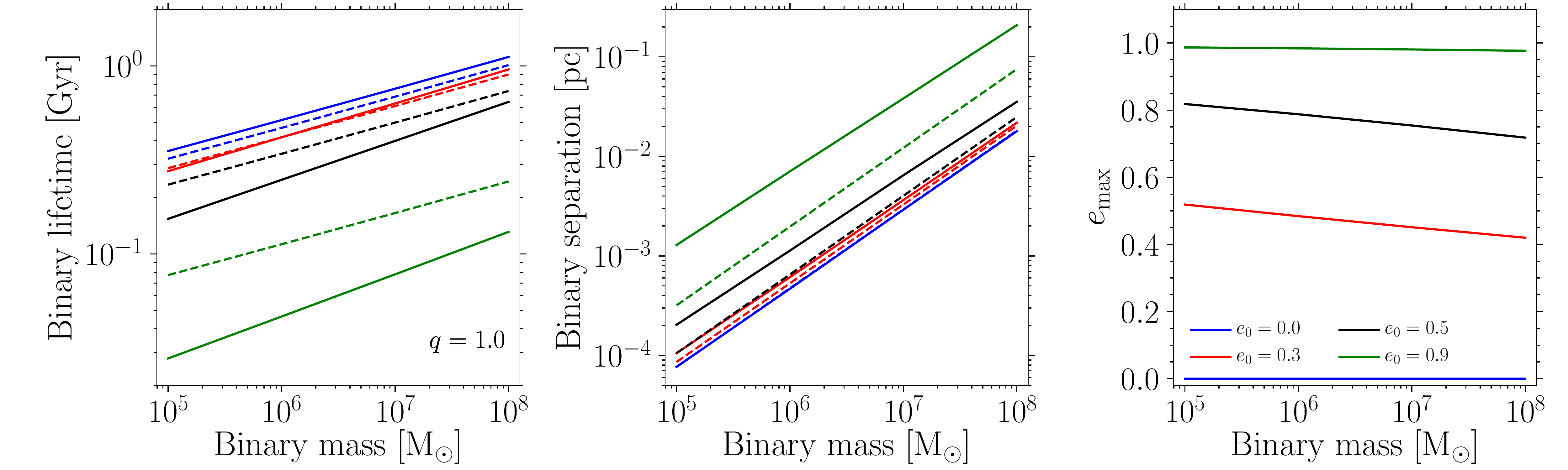}
\includegraphics[scale=0.35]{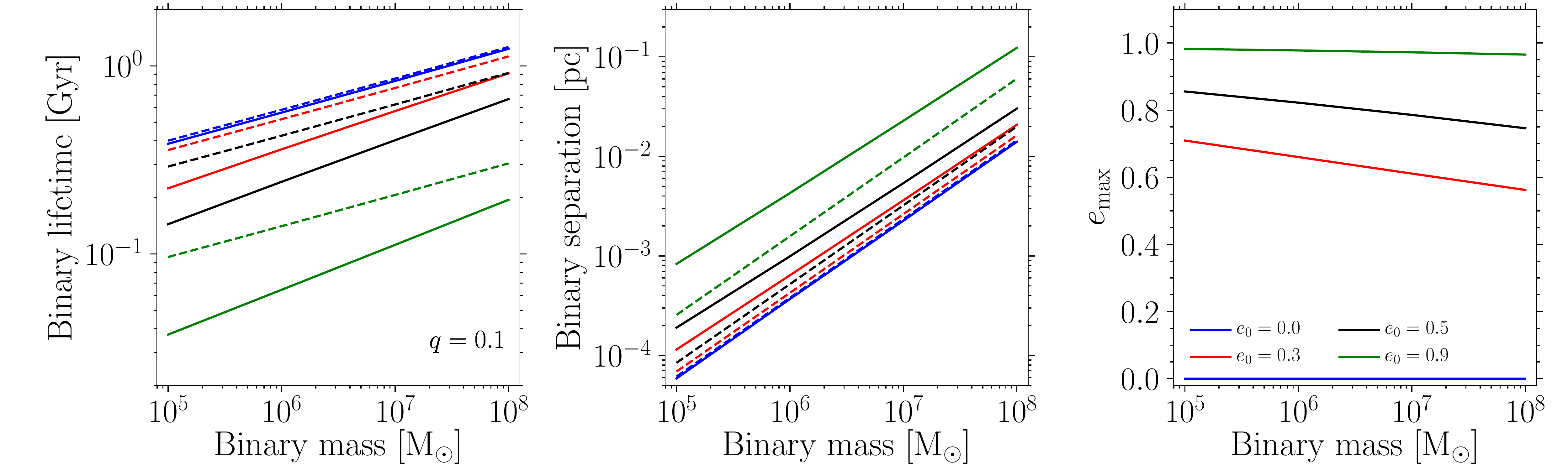}
\vspace{-7pt}
\caption{Lifetime $t(a_{\rm */gw})$ (left-hand panels), characteristic separation $a_{\rm */gw}$ (central panels), and maximum eccentricity reached $e_{\rm max}$ (right-hand panels) as a function of the binary total mass, considering four different initial eccentricities ($e_0 = 0.0, 0.3, 0.5$, and 0.9), for a mass ratio $q=1$ (top panels) and 0.1 (bottom panels). We remark that the mean eccentricity of the Keplerian binary at the end of the cuspy simulation in \citet{Tamfal2018} is $\sim$0.5. The density and velocity dispersion of the underlying host are computed using the Dehnen profile with index $\gamma=1$. In the left-hand and central panels, solid and dashed lines show the evolution of the binary's lifetime and characteristic separation when considering an evolving and fixed eccentricity, respectively. From the right-hand panels, it is evident that, at lower masses, binaries can acquire a higher eccentricity (if the eccentricity at pairing is already moderate) as a consequence of the longer interaction with the stellar background (except when $e_0 = 0.0$, for which $K \sim 0$). Therefore, black hole binaries of $10^{5}\msun$ have a shorter GW phase and, consequently, a reduced binary life-time.}
\vspace{-5pt}
\label{fig:sesana}
\end{figure*}
%%%%%%%% END FIGURE1

\section{Binary lifetime}\label{sec:binary_lifetime}

Consider the case of a black hole binary embedded in a stellar spherical background. The binary hardens via scattering off single stars plunging on nearly radial orbits, and
GW emission, which intervenes at the shortest distances.
The evolution of the semi-major axis $a$ is ruled by the two mechanisms and is given by the sum of two  terms which have different scalings with $a$ \citep{SesanaKhan2015}:

\begin{equation}
 \frac{da}{dt}=\frac{da}{dt}\biggl|_{\rm 3b} + \frac{da}{dt}\biggl|_{\rm gw} = -Aa^2 - \frac{B}{a^3},
 \label{eq:a_evolution}
\end{equation}

\noindent where the coefficients

\begin{equation}
 A=\frac{GH\rho_{\rm inf}}{\sigma_{\rm inf}}, \qquad B=\frac{64G^3 q M_{\rm BH,T}^3F(e)}{5c^5(1+q)^2}
\end{equation}

\noindent describe the three-body `star--black hole binary' close interaction and GW dissipation, respectively.
The binary has total  mass $M_{\rm BH,T}=M_{\rm BH,1} + M_{\rm BH,2}$ and mass ratio $q=M_{\rm BH,2}/M_{\rm BH,1}\le1$. $G$ is the gravitational constant and $c$ is the speed of light in vacuum. 
The coefficient  $A$ depends on the ratio between the stellar density $\rho_{\rm inf}$ and velocity dispersion $\sigma_{\rm inf}$ at the black hole binary
sphere of influence.  Thus, information of the mass of the binary is implicitly contained in the values of the density and velocity dispersion of the underlying 
stellar system. $H$ is a dimensionless hardening rate inferred from three-body scattering experiments, of the order of $\sim$15--20 \citep{Quinlan1996,Sesana2006}, which we set equal to 15.
The coefficient $B,$ related to the GW energy loss, is a sensitive  function of the binary eccentricity $e$, with the factor $F(e)=(1-e^2)^{-7/2}[1+(73/24)e^2+(37/96)e^4]$ \citep{Peters1963}.

Since the stellar hardening rate is $\propto a^2$ and the GW hardening rate is $\propto a^{-3}$,
binaries spend most of their time at the transition separation that can be estimated setting

\begin{equation}
\vert (da/dt)_{\rm 3b}|=|(da/dt)_{\rm gw}\vert .
\label{eq:equality}
\end{equation}

\noindent This occurs at a distance

\begin{equation}
 a_{\rm */gw} = \biggl[\frac{64G^2\sigma_{\rm inf}qM_{\rm BH,T}^3F(e)}{5c^5(1+q)^2H\rho_{\rm inf}}\biggr]^{1/5}
 \label{eq:a*gw}
\end{equation}

\noindent corresponding to a maximum in the hardening time-scale

\begin{equation}
 t(a_{\rm */gw})=\frac{\sigma_{\rm inf}}{GH\rho_{\rm inf}a_{\rm */gw}},
 \label{eq:t(a*gw)}
\end{equation}

\noindent which we refer to as binary lifetime in the following. 
$ t(a_{\rm */gw})$ displays a weak dependence on the mass ratio $q$, scaling as $q^{1/5},$ and a stronger dependence on the eccentricity, scaling as  $ t(a_{\rm */gw})\propto (1-e^2)^{7/10}$ (e.g. a reduction of a factor of 3.2 can be attained when the eccentricity is 0.9 compared to $e=0$).

Scattering experiments and direct $N$-body numerical simulations \citep[e.g.][]{Sesana2011,Khan2012,Khan2018a,Khan2018b} indicate that the binary eccentricity increases during the hardening phase by three-body scatterings. Thus, we couple Equation \eqref{eq:a_evolution} with the evolution equation for $e$, which accounts for the effects of stellar scatterings and GW cicularisation,

\begin{equation}
\frac{de}{dt}= a \dfrac{G \rho_{\rm inf} H K}{\sigma_{\rm inf}}  -\dfrac{304}{15} \dfrac{G^3 q M_{\rm BH,T}^3}{ c^5 (1+q)^2 a^4 (1-e^2)^{5/2}} \left(e+\dfrac{121}{304}e^3\right),
\label{eq:e_evolution}
\end{equation}

\noindent where the eccentricity growth rate $K$ is a numerical factor, calibrated against scattering experiments, which depends on $e$ and is of order $\sim$0 (for low values of $e$) and $\sim$0.2 otherwise  \citep{Quinlan1996,Sesana2006}.

Before proceeding, we need to estimate $\sigma_{\rm inf}$ and $\rho_{\rm inf}$. We do this for two different physical models.

\begin{figure*}
\vspace{-5pt}
\centering
\includegraphics[width=0.9\textwidth]{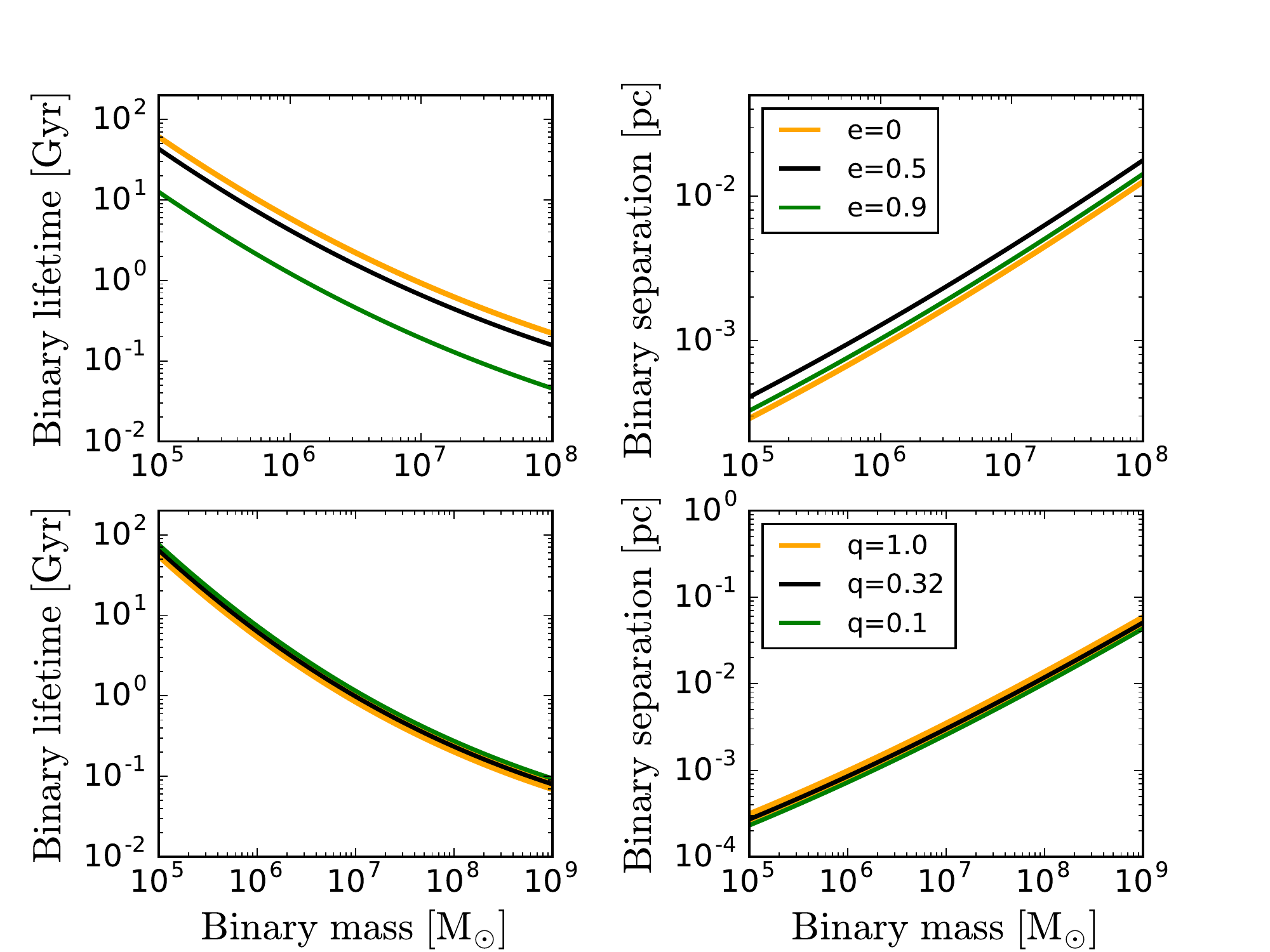}
\caption{Lifetime (left-hand panels) and characteristic separation (right-hand panels) of massive black hole binaries versus binary mass,
assuming the `black hole mass--S\'ersic index $n$' relation given by Equation~\eqref{eq:nS} for galaxies described by the Prugniel \&\ Simien profile, for (upper panels) equal mass ($q=1$) binaries  with 
constant, unevolved eccentricity $e = 0,0.5$, and 0.9 (orange, black, and green, respectively) and for (bottom panels) costant unevolved eccentricity $e=0$ with different mass ratio $q = 1.0,0.32$, and 0.1 (orange, black, and green, respectively). For the influence radius, we use here an average value of $r_{\rm inf}=0.023R_{\rm eff}$, deduced fitting the data of \citet{GrahamScott2015}.}
\vspace{-5pt}
\label{fig:all}
\end{figure*}

%%%%%%%%%%%
%Section 2.1%%%%
%%%%%%%%%%%

\subsection{Dehnen density profile}\label{sec:dehnen_density_profile}

\cite{SesanaKhan2015} calculated the coalescence time of a binary using as a reference model for the galaxy remnant the density profile described by \cite{Dehnen1993}:

\begin{equation}
 \rho(r) = \frac{(3-\gamma)M_{\rm sph,*}}{4\pi}\frac{r_0}{r^{\gamma}(r+r_0)^{(4-\gamma)}},
 \label{eq:dehnen}
\end{equation}

\noindent where $r_0$ is the scale radius and $0 \le \gamma < 3$ the inner logarithmic slope.
The velocity dispersion profile associated to this model is

\begin{equation}
 \sigma^2(r)=GM_{\rm sph,*}r^{\gamma}(r+r_0)^{4-\gamma}\int_{r}^{\infty}dr' \,\frac{r'^{(1-2\gamma)}}{(r'+r_0)^{(7-2\gamma)}}.
 \label{eq:Dehnen_sigma}
\end{equation}

The Dehnen profile \citep[see also][]{Tremaine1994} is an analytical model which is a generalisation of widely used models such as the \citeauthor{Hernquist1990} (\citeyear{Hernquist1990}; $\gamma = 1$) and \citeauthor{Jaffe1983} (\citeyear{Jaffe1983}; $\gamma = 2$) profiles. In projection, for $\gamma \sim 3/2$, it resembles quite well the de Vaucouleurs' law $R^{1/4}$  profile \citep{deVaucouleurs1948, deVaucouleurs1959}, 
which describes how the surface brightness of ellipticals and bulges of spirals  varies as a function of the projected distance $R$ from the centre.

To evaluate the density and velocity dispersion at the black hole binary sphere of influence $r=r_{\rm inf}$, we need to correlate the black hole binary mass with the stellar mass.
Under the assumption that scaling relations can be applied to the binary, when the merger is near completion, we proceed as follows:

\begin{itemize}

\item The stellar (bulge) mass $M_{\rm sph,*}$ is inferred from the $M_{\rm BH}$--$M_{\rm sph,*}$ relation taken from \cite{KormendyHo2013}: $M_{\rm BH,T}/10^9\rm{M_{\odot}} = 0.49(M_{\rm sph,*}/10^{11}\rm{M_{\odot}})^{1.16}.$

\item The effective radius $R_{\rm eff}$ is a function of the galaxy stellar mass and depends on the nature of the galaxy host. 
For massive ellipticals, bulges of spirals, and ultra-compact dwarfs, \citet{Dabringhausen2008} found $R_{\rm eff}/\rm{pc}=2.95(M_{\rm sph,*}/10^6\,\rm{M_{\odot}})^{0.596}.$

\item The scale radius $r_0$ is connected to the bulge effective radius $R_{\rm eff}$ through the relation $R_{\rm eff}\sim 0.75r_0(2^{1/(3-\gamma)}-1)^{-1}$ \citep{Dehnen1993}.

\item The influence radius, defined as the radius containing twice the binary mass in stars, 
i.e. $M_*(<r_{\rm inf})=2M_{\rm BH,T}$, is given by $r_{\rm inf}=r_0/\{[M_{\rm sph,*}/(2M_{\rm BH,T})]^{1/(3-\gamma)}-1\}$.

 \item The stellar density  $\rho_{\rm inf}$ is obtained from Equation~\eqref{eq:dehnen} at the influence radius.
 
 \item The velocity dispersion $\sigma_{\rm inf}$ is computed using Equation~\eqref{eq:Dehnen_sigma}.\footnote{In \cite{SesanaKhan2015}, $\sigma_{\rm inf}$ was computed using the $M_{\rm BH}$--$\sigma$
 correlation by \cite{KormendyHo2013}. We note, however, that this choice does not significantly change the results.}

\end{itemize}

\begin{figure}
\vspace{-12pt}
\centering
\includegraphics[width=0.5\textwidth]{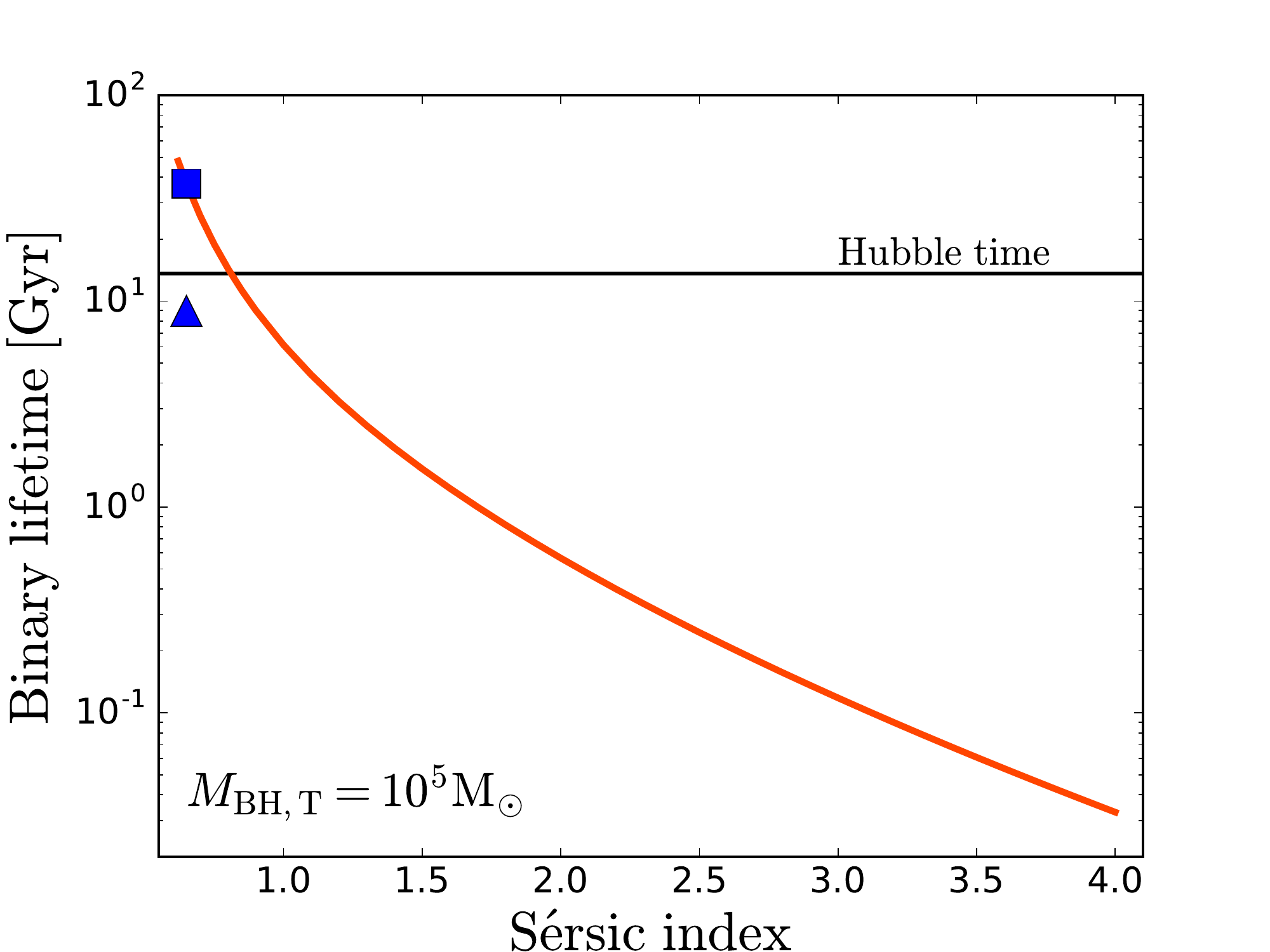}
\vspace{-5pt}
\caption{Binary lifetime versus S\'ersic index $n$, for a black hole binary of $10^5\msun$. 
The blue square indicates the time evaluated in this work binding the effective radius to the stellar mass of the galaxy with the relation found by \citet{Dabringhausen2008} and estimating the S\'ersic index from the \citet{Savorgnan2013} relation, using a costant eccentricity of $e = 0$. The blue triangle indicates the time derived assuming $e = 0.9$ (this value is close to $e_{\rm max}$ given in Figure~\ref{fig:sesana} for a binary of $10^5\msun$, when starting from $e_0 = 0.5$, the mean eccentricity at the end of the cuspy simulation in \citealt{Tamfal2018}). The horizontal line indicates the Hubble time.}
\vspace{-5pt}
\label{fig:t-Re}
\end{figure}

With the above prescriptions, we relate in a unique way $M_{\rm BH,T}$ to the properties of the host galaxy: $M_{\rm sph,*},r_0,R_{\rm eff},r_{\rm inf},\rho_{\rm inf}$, and $\sigma_{\rm inf}.$
This enables us to solve Equations~\eqref{eq:a_evolution} and \eqref{eq:e_evolution}, for different values of the initial eccentricity $e_0$ and an initial binary separation equal to the radius $a_0$ at which the enclosed stellar mass is equal to $2 M_{\rm BH,T}$  (when a Keplerian binary forms). We then compute the characteristic separation $a_{*/\rm gw}$ and the corresponding lifetime $t(a_{*/\rm gw})$ as a function of the binary mass, from Equations~\eqref{eq:a*gw} and \eqref {eq:t(a*gw)}, for a given value of $\gamma$, $q$, and $e_0$. This is illustrated in Figure~\ref{fig:sesana}, where we fix $\gamma = 1$ and consider two values of the mass ratio $q=0.1$ and 1 (we do not consider lower values of the mass ratio, as minor mergers with  $q\lesssim 0.1$ may lead to wandering black holes; \citealt{Callegari2008,Callegari2011,Colpi2014}). In Figure~\ref{fig:sesana}, $e_{\rm max}$ is the maximum binary eccentricity attained during evolution.

We note that lifetimes cover a wide range, from 30~Myr up to 1~Gyr,
show very weak dependence on $q$, and depend sensitively on $e_0$. LISA black hole binaries need to attain tiny separations compared to galactic scales, of the order of $10^{-3}$~pc ($e_0 = 0.9$) and $10^{-4}$~pc ($e_0 = 0.3$) for the lightest binaries of $10^5\msun$. These relations, computed using the expression of the velocity dispersion at the gravitational sphere of influence of the binary, are close to the ones computed by \cite{SesanaKhan2015}, who assumed non-evolving values of the eccentricity at the transition between stellar hardening and GW emission. We note that accounting for the eccentricity evolution reduces the binary lifetime by a factor of $\sim$2--10. Thus, in the following analysis, we treat the eccentricity as a constant and equal to zero, so that time-scales provide an upper limit. Note that, if the relevant radius at which the loss cone is kept full is $r_{\rm inf}={GM_{\rm BH,T}}{/\sigma^2(r_{\rm inf})}$, then the time-scales decrease by a factor of $\sim$2.

%%%%%%%%%%%
%Section 2.2%%%%
%%%%%%%%%%%

\begin{figure*}
\vspace{-11pt}
\centering
\includegraphics[width=0.85\textwidth]{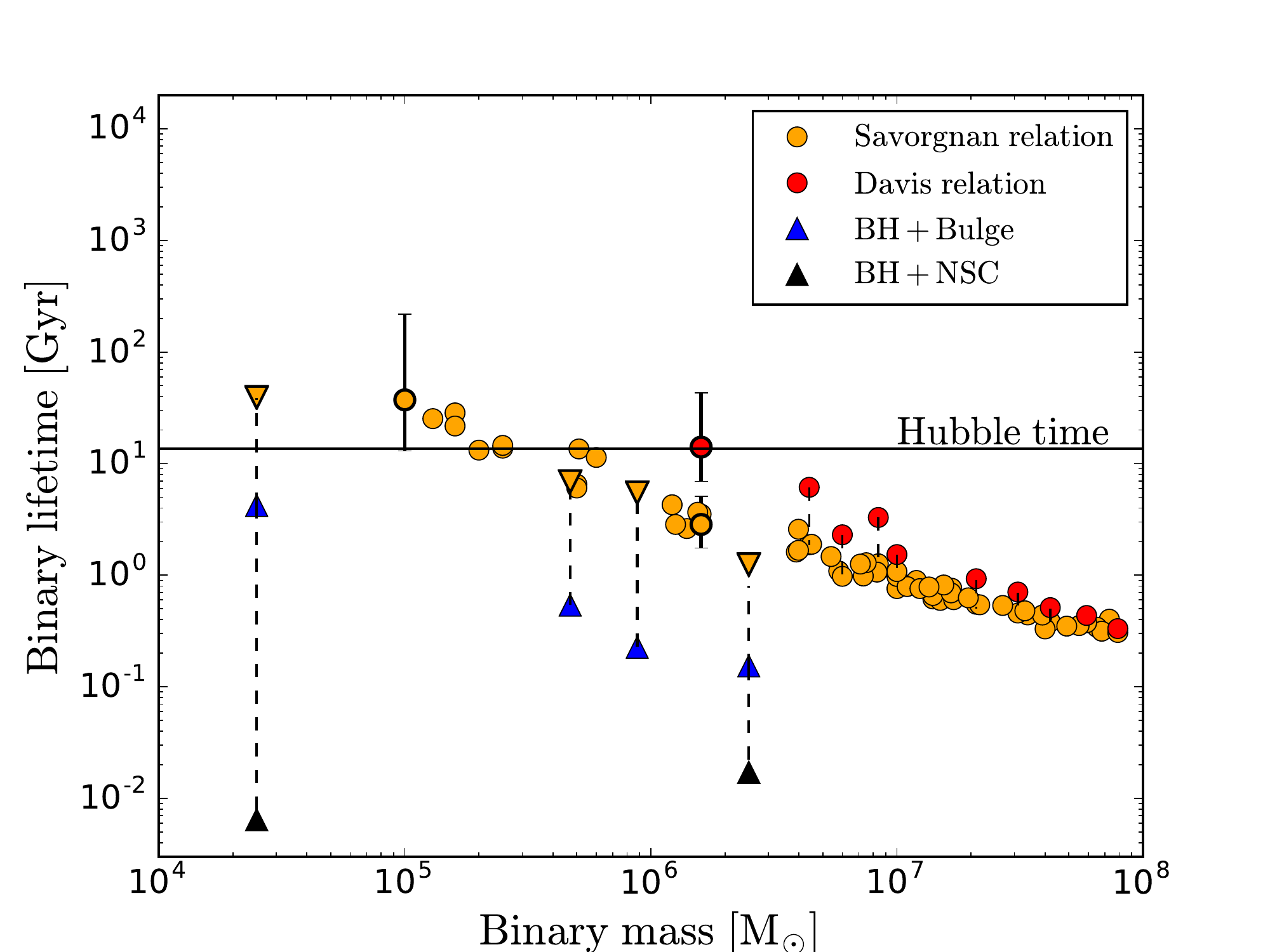}
\caption{Binary lifetime versus binary mass for the sample of S\'ersic galaxies from \citet{GrahamScott2015}, inferred using the Prugniel \& Simien stellar density profile, and correlating the binary mass with the  S\'ersic index $n$, adopting both Equation~\eqref{eq:nS} \citep{Savorgnan2013} -- orange dots -- and \eqref{eq:nD}  \citep{Davis18} -- red dots. For two values of the binary mass ($10^5$ and $1.6 \times 10^6$~M$_{\odot}$), we also show the error associated to the uncertainties of the relations described by Equations~\eqref{eq:nS} -- for both values of the masses -- and \eqref{eq:nD} -- for $1.6 \times 10^6$~M$_{\odot}$ only. We additionally consider four black holes studied in \citet{Nguyen2018}, for which we calculate lifetimes using either the relations given in Section~\ref{sec:prugniel_and_simien_density_profile} [Equation~\eqref{eq:nS}; orange triangles] or the data given in the original article for a bulge decomposition of the host galaxy (blue triangles). These galaxies host a nuclear star cluster (NSC) in their centre. Therefore, we estimate the lifetimes considering the NSC decomposition (black triangles). In this Figure, the binary has a mass ratio $q=1$ and eccentricity $e=0$, kept constant. Thus, the plot provides upper limits to the binary lifetime. The Hubble time is indicated by a horizontal line.
}
\vspace{-5pt}
\label{fig:Ser}
\end{figure*}

\begin{figure*}
\centering
\vspace{-11pt}
\includegraphics[width=0.85\textwidth]{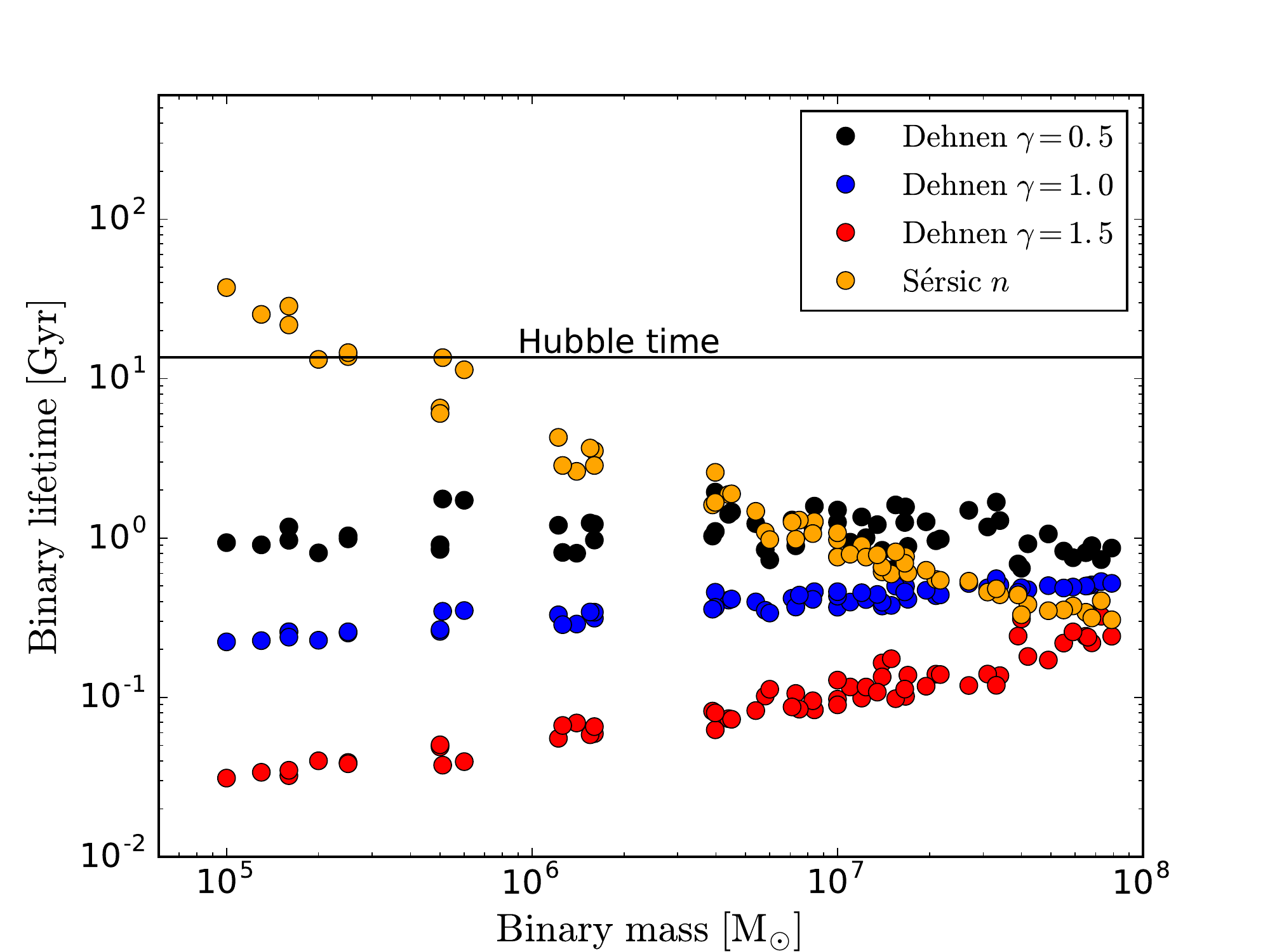}
\caption{Binary lifetimes versus binary mass for the sample of S\'ersic galaxies taken from \citet{GrahamScott2015}, using the Prugniel \& Simien stellar density profile in orange as in Figure~\ref{fig:Ser}. For a comparison, black, blue, and red dots refer to the same data-set but computing the binary lifetime $t_{*/\rm gw}$  using Dehnen profiles with $\gamma=0.5,1.0$, and 1.5, respectively. For homogeneity in the comparison, we computed here the black holes sphere of influence in terms of the stellar velocity dispersion. The Hubble time is indicated by a horizontal line.}
\vspace{-5pt}
\label{fig:comp}
\end{figure*}

\subsection{Prugniel and Simien density profile}\label{sec:prugniel_and_simien_density_profile}

Galaxies in the \cite{GrahamScott2015} data set are divided in S\'ersic and core-S\'ersic galaxies.
The term \emph{S\'ersic galaxy} is used to denote galaxies (ellipticals and bulges of disc galaxies) whose surface brightness  $I$
is described by the S\'ersic model \citep{Sersic1963, Sersic1968},   $I\propto R^{1/n}$.
The term \emph{Core-S\'ersic galaxy} refers to a galaxy whose main spheroidal component has a partially depleted core 
(i.e. a central stellar deficit of light that is not due to dust, enclosed in a radius $R \sim 0.01R_{\rm eff}$) such that the surface brightness profile is well described by the core-S\'ersic model \citep{Graham2003}, which joins a single inner power-law profile to an outer S\'ersic profile.
\cite{ScottGraham2013} have found that the $M_{\rm BH}$--$M_{\rm sph,*}$  relation is different for this type of galaxies: core-S\'ersic galaxies follow a linear relation, 
whereas S\'ersic galaxies a quadratic one. The bend in the relation occurs at $M_{\rm BH} \sim 2\times 10^8\msun$, with S\'ersic galaxies hosting the smallest black holes.

We calculate again the binary lifetime considering only S\'ersic galaxies, which host central black holes in the range of interest of LISA.
This time, however, we use the density profile developed by \cite{PrugnielSimien1997} to describe S\'ersic galaxies \citep[a complete analysis of the model is reported in][]{TerzicGraham2005}.
Density as a function of radius $r$ is given by

\begin{equation}
 \rho(r) =\rho_0 \biggl(\frac{r}{R_{\rm eff}}\biggr)^{-p} e^{-b(r/R_{\rm eff})^{1/n}},
 \label{eq:rho-n}
\end{equation}

\noindent which depends on the curvature of the profile $n$, the normalization factor $\rho_0$, and the effective radius $R_{\rm eff}$. The parameters  $p$ and $b$ are a function of $n,$
as described below.
The velocity dispersion is given by

\begin{equation}
 \sigma^2(r) 
 = \frac{4\pi G\rho_0^2R_{\rm eff}^2n^2b^{2n(p-1)}}{\rho(r)}\\ \int_{Z}^{\infty}\bar{Z}^{-n(p+1)-1}e^{-\bar{Z}}\gamma[n(3-p),\bar{Z}]d\bar{Z} 
\end{equation}

\noindent where $Z=b(r/R_{\rm eff})^{1/n}$, and here $\gamma$ is the lower incomplete Gamma function.

The parameters used in these equations are calculated from the following relations:
\begin{itemize}

\item The stellar mass of the spheroid is inferred either from \citep{ScottGraham2013}
 
 \begin{equation}
 M_{\rm BH,T}/10^7\rm{M_{\odot}} = 7.89\,(M_{\rm sph,*}/2\times 10^{10} \rm{M_{\odot}})^{2.22},
 \label{eq:M-sersic}
 \end{equation}
 
 \noindent when evaluating the trends of Figure~\ref{fig:all}, or directly from the sample of \citet{GrahamScott2015}, when plotting the data-points of Figure~\ref{fig:Ser}.
 
  \item $R_{\rm eff}$ is connected to the stellar mass through the relation 
 $R_{\rm eff}/{\rm pc}=2.95(M_{\rm sph,*}/10^6\,\rm{M_{\odot}})^{0.596}$ for the bulges of spirals and ultra-compact dwarfs \citep{Dabringhausen2008}.

 \item The index \emph{n} is connected to the black hole mass through the relation taken from \cite{Savorgnan2013} for S\'ersic bulges:
 
 \begin{equation}
 \log_{10}(M_{\rm BH,T}/\msun)=7.73+4.11\log_{10}(n/3). 
 \label{eq:nS}
 \end{equation}

\item $p$ is connected to $n$ by $p=1.0-0.6097/n+0.05563/n^2$, for $0.6<n<10$ and $10^{-2}\le r/R_{\rm eff}\le10^3$ \citep{Marquez2000}.
 This relation is obtained through a high-quality match between the exact, de-projected S\'ersic profiles (solved numerically) and the above expression of $\rho(r)$.
 
 \item $b$ is chosen to ensure that $R_{\rm eff}$ contains half the (projected) galaxy light and is obtained by solving the equation $\Gamma(2n)=2\gamma(2n,b)$, where $\Gamma$ and $\gamma$ are the Gamma and lower incomplete Gamma function, respectively.
 A good approximation of $b$ for $0.5<n<10$ is $b=2n-1/3+0.009876/n$ \citep{PrugnielSimien1997}.
 
 \item The density $\rho_0$ is inferred from Equation~\eqref{eq:rho-n} through integration over the mass distribution.
 
 \item The binary influence radius $r_{\rm inf}$ has been deduced for every galaxy in the sample from the relation $r_{\rm inf}={GM_{\rm BH,T}}{/\sigma^2(r_{\rm inf})}.$

\end{itemize}

In Figure~\ref{fig:all}, we plot the binary lifetime and separation as a function of $M_{\rm BH,T}$
for different values of (unevolved) eccentricity and mass ratio. In S\'ersic galaxy models,  the binary lifetime increases with decreasing binary mass, in contrast to the Dehnen model (Figure~\ref{fig:sesana}). This trend mirrors the correlation between the black hole mass and the index $n$, as lower-mass black holes are hosted in  galaxies with lower values of $n$. The binary lifetime is proportional to $(\sigma_{\rm inf}/\rho_{\rm inf})^{4/5}$ [see Equations~\eqref{eq:a*gw} and \eqref{eq:t(a*gw)}] which, in Prugniel \& Simien models with low $n$, increases with decreasing $n$. We find that  the transition separation $a_{\rm */gw}$ falls in a similar way to that of the Dehnen models.

Figure~\ref{fig:t-Re} shows the lifetime of a binary black hole of $10^5\msun$ in a host galaxy of $1.3\times 10^9\msun$ as a function of the S\'ersic index $n,$  to illustrate the dependence of this time-scale on this index, which determines both the density and dispersion velocity at the binary black hole influence radius.

%%%%%%%%%%
%Section 3%%%%%
%%%%%%%%%%%

\section{Lifetimes: black hole mass versus S\'ersic index}\label{sec:Nguyen}

In this section, we show the strong dependence of the binary lifetimes on the S\'ersic index. In Figure~\ref{fig:Ser}, we plot with orange dots the binary lifetime calculated for the black holes in the \cite{GrahamScott2015} sample (for the case $q=1, e=0$) using Equation~\eqref{eq:nS}.
Lifetimes range between 0.3~Gyr and the Hubble time for black hole binaries with masses above a few $10^5\msun$, but exceed the Hubble time for lighter binaries. These lifetimes are upper limits, as binaries are expected to develop large eccentricities during the hardening which reduce the value by factors of $\sim$2--5 (see Figure~\ref{fig:all}).

Recently, \cite{Davis18} proposed a revised version of the $M_{\rm BH}$--$n$ relation:
 
 \begin{equation}
     \log_{10}(M_{\rm BH,T}/\msun)=7.45+2.76\log_{10}(n/2.20).
     \label{eq:nD}
 \end{equation}
 This new relation gives lower values of $n$ for a given (low) black hole mass, compared to the relation by \cite{Savorgnan2013}. 
 Red dots in Figure~\ref{fig:Ser} 
indicate binary lifetimes inferred using Equation~\eqref{eq:nD}, for a few select black holes.

Both relations have a very large scatter, and we computed the uncertainty on $t(a_{*,\rm gw})$ for two select black hole mass ($10^5$ and $1.6 \times 10^6$~M$_{\odot}$) highlighted in Figure~\ref{fig:Ser} with black circles and associated errors. For example, using the Davis et al. (Savorgnan et al.) relation and scatter, the uncertainty on $t(a_{*,\rm gw})$ when $M_{\rm BH,T} = 1.6 \times 10^6$~M$_{\odot}$ is of a factor of 7--8 (of 2). For $M_{\rm BH,T} = 10^5\msun$, the scatter is $\sim$10 using Savorgnan et al.

 \citet{Nguyen2018} published a high-resolution study of four black holes hosted in nearby low-mass early-type galactic nuclei (M32, NGC~205, NGC~5102, and NGC~5206), providing the values of the black hole mass, host galaxy stellar mass, S\'{e}rsic index and effective radius of the bulge (and of the NSC, when possible), and black hole influence radius. We computed the expected binary lifetimes using the published values of $M_{\rm BH}$, $n$, $R_{\rm eff}, r_{\rm inf}$ and the Prugniel \& Simien model for the density and stellar dispersion at $r_{\rm inf}$, and compared them with the ones we inferred using Equation~\eqref{eq:nS}. In Figure~\ref{fig:Ser}, we show how the binary lifetime decreases by one or more orders of magnitude when, instead of using the relations in Section~\ref{sec:prugniel_and_simien_density_profile} (orange triangles), we use the data given in \citet{Nguyen2018} of the bulge (blue triangles) or the NSC (black triangles).

This illustrates how sensitive is the dependence of the binary lifetimes on the underlying properties of the stellar background.

%%%%%%%%%%%
%Section 4%%%%
%%%%%%%%%%%

\section{Dehnen versus Prugniel \& Simien}\label{sec:model_comparison}

It is interesting to carry out a direct comparison between the  two models. To this purpose, we calculate the binary lifetime for the galaxies of the \citet{GrahamScott2015} sample,
using both the Denhen profile (varying $\gamma$), and the Prugniel \& Simien model along the $M_{\rm BH}$--$n$ correlation of Equation~\eqref{eq:nS}.
For consistency, we define the binary influence radius as $r_{\rm inf}={GM_{\rm BH,T}}{/\sigma^2(r_{\rm inf})}$ for both models.
Figure~\ref{fig:comp} gives in black, blue, and red the time-scales corresponding to the Dehnen slopes $\gamma=0.5$, 1.0, and 1.5, respectively. In orange,
we give the result of the  Prugniel $\&$ Simien model to contrast the different trends of $t(a_{*/\rm gw})$ with the black hole binary mass.

\begin{figure}
\vspace{-10pt}
\centering
\includegraphics[width=0.5\textwidth]{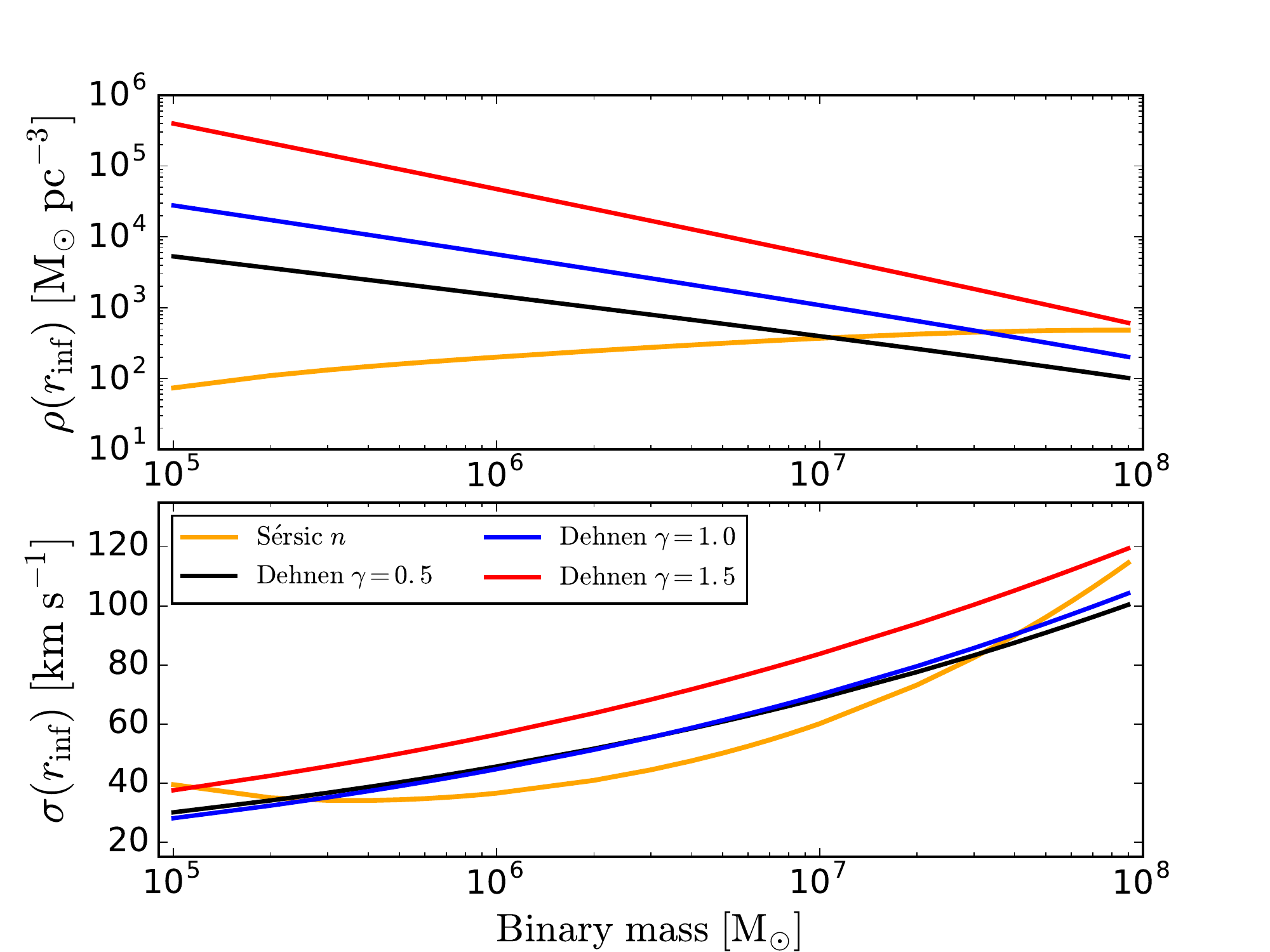}
\vspace{-5pt}
\caption{Upper panel: stellar density evaluated at the binary gravitational influence radius, $\rho(r_{\rm inf})$,
versus binary mass. Bottom panel: same for the stellar velocity dispersion, $\sigma(r_{\rm inf})$. The orange lines refer to the Prugniel \& Simien profile, along the $M_{\rm BH}$--$n$ correlation [Equation~\eqref{eq:nS}], whereas the
red, blue, and black lines refer to the Dehnen profile with $\gamma=1.5,1.0$, and 0.5, respectively.
}
\vspace{-5pt}
\label{fig:rho-vel}
\end{figure}

This difference can be understood by comparing, amongst the different models, the values of the density and velocity dispersion at the binary gravitational influence radius
as highlighted in Figure~\ref{fig:rho-vel}.  The dependence of both density and velocity dispersion on $M_{\rm BH,T}$ 
comes from the correlations $M_{\rm BH}$--$M_{\rm sph,*}$, $M_{\rm BH}$--$n$, and $R_{\rm eff}$--$M_{\rm sph,*}$, and from $r_{\rm inf}$ itself.
We note a remarkable difference among the values of the densities $\rho(r_{\rm inf})$ between the Dehnen and the Prugniel \& Simien models.
In all Dehnen profiles, the stellar density is a decreasing function of the black hole binary mass and can reach values as high as $\sim$$10^5\,\msun\,\rm pc^{-3}$ for $M_{\rm BH,T} = 10^5$~M$_{\odot}$ and $\gamma = 1.5$. Moreover, the change in stellar density is more pronounced for higher values of $\gamma$. By contrast, in S\'ersic galaxies, the run of $\rho(r_{\rm inf})$ versus $M_{\rm BH,T}$ is the opposite, with remarkably lower values, down to about $\sim$$100\,\msun\,\rm pc^{-3}$ for $M_{\rm BH,T} = 10^5$~M$_{\odot}$. The density changes by a factor of $\sim$10 over the entire range of black hole binary masses. There is no noticeable difference instead in the values of the stellar velocity dispersion  between the two families of models.

%%%%%%%%%%%
%Section 5%%%%%
%%%%%%%%%%%

\section{Conclusions}\label{sec:conclusions}

In this paper, we carried on an estimate of the hardening time of black hole binaries in the mass interval between $10^5$ and $10^8\msun$, relevant for LISA \citep{Amaro-Seoane2017}, under the assumption that binary contraction is driven by individual scatterings off stars. To this aim, we employed the empirical relation for the binary lifetime described in \cite{SesanaKhan2015}, assuming that the black holes inhabit the stellar, central environment of local galaxies. In the analysis, we used two different models for the underlying stellar density and velocity dispersion profiles: the \cite{Dehnen1993} model and the \cite{PrugnielSimien1997} model, the latter describing galaxies with surface brightness fit by the S\'ersic profile. Using local observed correlations between the black hole mass $M_{\rm BH}$ and the mass of the spheroid $M_{\rm sph,*}$ \citep{KormendyHo2013, ScottGraham2013}, and the S\'ersic index $n$ \citep{Savorgnan2013, Davis18}  of the underlying host galaxy, we find different trends and a large spread of the binary lifetimes. 

In S\'ersic galaxies, described by a Prugniel \& Simien profile, binary lifetimes increase with decreasing mass and exceed the Hubble time below $\sim$ $5 \times 10^5\msun$. These time-scales have been computed assuming null eccentricity, thus providing upper limits. In high-resolution simulations of isolated collisions
between dwarf galaxies, \cite{Tamfal2018} find that, when a binary forms, it already carries a rather large eccentricity, with mean $e \sim 0.5$. Furthermore, we know that binaries increase their eccentricity during the hardening phase by stellar scattering. Thus, when considering the eccentricity evolution, lifetimes are comparable to/smaller than the Hubble time. We further caution that uncertainties in the observed correlations (which are larger in the low-mass tail of the observed black hole population) give a spread of about one order of magnitude. 

If Dehnen profiles represent the underlying stellar population of galaxies hosting black holes, we find that binary lifetimes display a nearly constant or decreasing trend with decreasing black hole binary mass.
Depending on the slope of the Dehnen profile, lifetimes are overall in the range between 30~Myr and 2~Gyr.
If Dehnen profiles and high S\'ersic indices are representative only of the high-mass tail ($\gtrsim 10^7\msun$) of the black hole 
population, the black holes for which LISA will be most sensitive are subject to time delays that can span a wide range, 
depending on how steeply the stellar profiles rise within their gravitational sphere of influence.

 Recent high-resolution observations of four strongly nucleated dwarf elliptical galaxies which host a central black hole \citep{Nguyen2018} provided S\'ersic indices for the bulge component systematically larger ($n>1.4$ for the bulge component) than those expected from the correlations by \cite {Savorgnan2013} and \cite{Davis18}, yielding lifetimes even lower than 1~Gyr.
 For two galaxies (M32 and NGC~205), there is also clear evidence of a central NSC with a mass in excess of that of the central black hole, a S\'ersic index larger than that of the bulge, and an effective radius larger than the black hole
 gravitational sphere of influence \citep{Nguyen2018}. Under these hypotheses, the inferred lifetimes are of the order of 10~Myr.
 In the sample from \citet{Graham2003} and \citet{GrahamScott2015}, the black holes with mass $\sim$$10^6\msun$ are hosted in galaxies
with stellar masses in the range between a few $10^8\msun$ and $\sim$$10^{10}\msun$, highlighting the high degree of dispersion of the
black hole--stellar mass correlation. However, we note that it is in this mass range that galaxies in cluster environments such as Virgo, Coma, and Fornax \citep{Sanchez18} host NSCs. The nucleation fraction has its peak of 90 per cent 
around a mass of $M_*\sim 10^9\msun$, and declines at lower and higher stellar masses. Thus, we speculate that accounting for
the presence of a NSC during the hardening of a black hole binary might lead to (i) shorter lifetimes and (ii) a turn over in the 
trend just around a pivotal mass of $10^6\msun$. Numerical simulations of black hole binary hardening in nucleated galaxies have never been explored in the literature, so far, and work is in progress along this line.  At galaxy masses below $10^8\msun$, NSCs are not any longer ubiquitous \citep[as shown in fig.~2 of][]{Sanchez18} and observations of dwarfs show that their inner logarithmic slopes range widely 
\citep{Gebhardt96,Glass2011,Geha17,Munoz18}, with a tendency to have steeper
profiles out to magnitudes of at least about -16 \citep{Gebhardt96}. Entering the realm of dwarf galaxies makes extrapolations 
of scaling relations and dynamics extremely uncertain and troublesome.

Only a small fraction of dwarf spheroidals
or galaxies of similar mass are however expected to host central black holes \citep[][]{VanWassenhove2010}. For black holes hosted in low-mass galaxies with cored profiles and dark matter-dominated, such
as dwarf spheroidals, a bottleneck exists along the path to coalescence in the early stages of the
merger, ruled by dynamical friction \citep[][]{Read2006}. The drag exerted on the
black hole mainly by the dark matter background is not effective enough and this
leads to a failure (or major delay) in the formation of a binary, which is the phase
anticipating that of stellar hardening \citep[][]{Tamfal2018}. Thus, LISA coalescence
events with binary black holes of a few $10^5 \Msun$ require dense environments.
 
We remind the reader that our results rely on observations (e.g. scaling relations) of the local universe. Thus, our estimates of the lifetimes depict hypothetical mergers at $z \lesssim 1$. At higher redshifts ($1\lesssim z\lesssim 
3$), galaxies show evolution in their averaged size (at fixed stellar galaxy mass) with $R_{\rm eff}$ scaling as $(1+z)^{-\alpha}$, with $\alpha$ varying from 0.7 (for late-type galaxies) to 1.48 (for early-type galaxies) \citep{vanderwel14}. Thus, at $z\sim 3$, a factor of $\sim $4 in the reduction of the  effective radius can lead to an increase of $\sim$64 in the density, if the galaxy model can be scaled self-similarly. Indeed, one recent study on massive galaxy mergers \citep{Khan2016} showed the hardening time of two massive black holes to decrease from $\gtrsim$10~Gyr at $z = 0$ to $\sim$ $10^7$~yr at $z = 3.3$ (see also the discussion in \citealt{Mayer17}). We expect the trend to be in the same direction, and there is the need to enlarge the sample of high-resolution simulations to consolidate this argument. However, one should remain aware that also simulations have uncertainties and results, especially for very local quantities,
are dependent on sub-grid physics and resolution.

Mergers of LISA black holes at very high redshift ($z \sim 10$) are even more difficult to model, as cosmological simulations are 
hampered by lack of resolution on the low-mass scales involved in this process \citep{Tremmel2017}. 
The black hole dynamics is by far more complex to describe, as halos are non-relaxed systems and subjected to repeated, multiple  interactions. Gas clumps can make the dynamics stochastic, and star formation and feedback can change the underlying background, broadening significantly the lifetime distribution for the binaries in the LISA-relevant mass range \citep{Bellovary2018,Pfister19}. Delay times between halo-halo mergers and black hole mergers can be computed using semi-analytical models which include also the formation of triple systems as a vehicle for the formation and coalescence of these light black holes \citep{Bonetti2018b}.

\bibliographystyle{mnras}
\bibliography{lifetime_of_BBHs_in_Sersic_galaxies}

% Don't change these lines
\bsp	% typesetting comment
\label{lastpage}
\end{document}